\documentclass[12pt]{article}

\usepackage[totalheight = 23cm, totalwidth = 17cm]{geometry}
\usepackage{amssymb,amsmath,amsfonts,amsbsy,graphicx}
\usepackage[colorlinks=true,linkcolor=blue,citecolor=blue]{hyperref}
\usepackage{subfigure,ulem,cite}

\newcommand{\mpl}{m_{\rm Pl}}
\newcommand{\calO}{{\cal O}}

\newcommand{\calR}{{\cal R}}

\begin{document}

\begin{titlepage}

\begin{center}

\hfill APCTP Pre2017-007\\

\vskip .75in

{\LARGE \bf Cosmological stochastic Higgs stabilization}

\vskip .75in

{\large Jinn-Ouk Gong$^{a,b}$ and Naoya Kitajima$^{a,c}$}

\vskip 0.25in

\textit{
   ${}^{a}$Asia Pacific Center for Theoretical Physics, Pohang 37673, Korea
   \\
   ${}^{b}$Department of Physics, Postech, Pohang 37673, Korea
   \\
   ${}^{c}$Department of Physics, Nagoya University, Nagoya 464-8602, Japan
}

\end{center}
\vskip .5in

\begin{abstract}

We show that the stochastic evolution of an interacting system of the Higgs and a spectator scalar field naturally gives rise to an enhanced probability of settling down at the electroweak vacuum at the end of inflation. Subsequent destabilization due to parametric resonance between the Higgs and the spectator field can be avoided in a wide parameter range. We further argue that the spectator field can play the role of dark matter.

\end{abstract}

\end{titlepage}

\section{Introduction}

%ew vacuum is unstable

The mass of the Higgs measured at the Large Hadron Collider, $m_h = 125.09 \pm 0.21 \pm 0.11$ GeV~\cite{Aad:2015zhl}, along with the top quark mass $m_t = 173.34 \pm 0.76 \pm 0.3$ GeV~\cite{ATLAS:2014wva}, indicates that the effective Higgs quartic coupling $\lambda_{\rm eff}(h)$ for $h \gg v \approx 246$ GeV, with the effective potential 
\begin{equation}
V(h) = \frac{\lambda_{\rm eff}(h)}{4}h^4 \, ,
\end{equation}
becomes negative at an energy scale $\Lambda = 10^{10}$ -- $10^{11}$ GeV, much bigger than the electroweak scale~\cite{Degrassi:2012ry,Buttazzo:2013uya} if the standard model is valid up to such high energy scales. Thus beyond $\Lambda$ the Higgs potential develops a true minimum with a large negative energy density, implying that the electroweak vacuum is only metastable. Although absolute stability is excluded at 99\% confidence level, the lifetime for the tunneling from the electroweak vacuum to the true vacuum at large field values is longer than the age of the universe~\cite{EliasMiro:2011aa}.

%higgs is not stabilized in the early universe

However, the cosmological evolution of the Higgs during inflation in the early universe makes the possibility of dwelling in the electroweak vacuum very unlikely \cite{Espinosa:2007qp,Kobakhidze:2013tn,Kearney:2015vba,Espinosa:2015qea,Hook:2014uia}. If the Higgs is only coupled to the standard model particle species, it can be treated as an effectively massless scalar field during inflation. Then the Higgs acquires quantum fluctuations of $\calO(H_{\rm inf})$ with $H_{\rm inf}$ being the Hubble parameter during inflation. The current bound on the tensor-to-scalar ratio $r_{0.05} < 0.07$ at 95\% confidence level~\cite{Array:2015xqh} gives
\begin{equation}
H_{\rm inf} \lesssim A_\calR\pi\sqrt{\frac{r}{2}} \mpl
\approx 6.73 \times 10^{13} {\rm GeV} \left( \frac{r}{0.07} \right)^{1/2} \, ,
\end{equation}
where $A_\calR \approx 2.21 \times 10^{-9}$ is the amplitude of the power spectrum of the primordial curvature perturbation at $k = 0.05/{\rm Mpc}$~\cite{Ade:2015lrj} and $\mpl \approx 2.43 \times 10^{18}$ GeV is the reduced Planck mass. Thus, even if initially at the electroweak vacuum near the origin, after inflation the Higgs is very likely to be located in the unstable region due to large quantum fluctuations of $\delta h = {\cal O}(H_{\rm inf}) \gg \Lambda$. A number of solutions to this problem has been suggested, e.g. by introducing a non-minimal coupling to gravity~\cite{Espinosa:2007qp,Herranen:2014cua}, a direct coupling to the inflaton~\cite{Lebedev:2012sy} or both~\cite{Ema:2017loe}, a finite temperature effect~\cite{Fairbairn:2014zia}, or a Hubble-induced mass~\cite{Kamada:2014ufa}.

%we present a novel scenario

In this article, we present a novel aspect for stabilizing the Higgs during inflation. Introducing a (gauge-singlet) scalar spectator field $S$ that couples to the Higgs as $\lambda_{hS}h^2S^2$, we find that the stochastic dynamics with this quartic interaction gives rise to a strongly enhanced probability of settling down at the electroweak vacuum at the end of inflation. The stability of the Higgs in the presence of a coupled singlet has been studied before, but we do not require a significant change of the renormalization group equations~\cite{Gonderinger:2009jp,Khan:2014kba}, a large vacuum expectation value of the singlet~\cite{Lebedev:2012zw,EliasMiro:2012ay} or Planck-suppressed interaction~\cite{Ema:2016ehh} to make the Higgs potential absolutely stable. Furthermore, even after possible parametric resonance after inflation we still have a good chance of stable electroweak vacuum 
compared to the case with the Higgs-inflaton and non-minimal couplings~\cite{Ema:2016kpf,Kohri:2016wof,Enqvist:2016mqj,Postma:2017hbk}.
We also find that for a stable singlet whose lifetime is longer than the age of the universe, depending on reasonable choices of the parameters in the Lagrangian, the singlet scalar whose interaction with the Higgs stabilizes the electroweak vacuum can also play the role of dark matter~\cite{McDonald:1993ex,Burgess:2000yq}.

\section{Multi-field stochastic dynamics during inflation}

Let us consider a singlet extension of the standard model with the following potential for the Higgs field $h$ and the singlet spectator field $S$:
\begin{equation}
V(h,S) = \frac{1}{4}\lambda_{\rm eff}(h) h^4 + \frac{1}{2}m_S^2S^2 
+ \frac{1}{2}\lambda_{hS} h^2S^2 + \frac{1}{4} \lambda_S S^4 \, .
\end{equation}
Precisely speaking, $\lambda_S$ and $\lambda_{hS}$ also depend on the energy scale. However, they do not change drastically and we approximately set them constant. In what follows, we use scale-independent values for $\lambda_S$ and $\lambda_{hS}$. In addition, we assume the singlet mass $m_S$ is much smaller than $H_{\rm inf}$ and neglect the second term in our analysis.

Throughout this article, we focus on relatively large inflation scale where the quantum fluctuation is a dominant force to drive the Higgs field out of the metastable region. Then we adopt stochastic approach via Fokker-Planck equation and neglect tunneling by Coleman-de Luccia~\cite{Coleman:1980aw} and Hawking-Moss instanton~\cite{Hawking:1981fz}. Under the slow-roll approximation, a slowly-varying course-grained field $\phi = \{h,S\}$ is described by stochastic classical theory with the Langevin equation including a stochastic Gaussian noise. Then the evolution of the probability distribution function $P(h,S,N)$ for the Higgs and the spectator to have values $h$ and $S$ respectively at the $e$-folding number $N$ is described by the Fokker-Planck equation~\cite{Mollerach:1990zf}:
\begin{equation}
\frac{\partial{P(h,S,N)}}{\partial{N}} = 
\sum_{\phi = h,S} \frac{\partial}{\partial\phi} 
\left[ \frac{V_\phi P(h,S,N)}{3H^2} + 
\frac{H^2}{8\pi^2} \frac{\partial P(h,S,N)}{\partial\phi} \right] \, .
\label{eq:FPE}
\end{equation}
Note that cross-derivative terms are absent because the quantum fluctuation for each field is independent. In what follows, we assume for simplicity that during inflation the Hubble parameter takes a constant value, $H_{\rm inf}$. Also, for computational simplicity, we set the initial conditions for the probability distribution as $P(h,S,0) = \delta(h)\delta(S)$, i.e. we assume that both the Higgs and the singlet field are placed at the origin initially. Note however that the distribution at later times is not sensitive to the initial distribution because of large effective masses for both fields $\sim \sqrt{\lambda_{hS}}H_{\rm inf}$, and the mean field values quickly approach to zero.

Eq.~\eqref{eq:FPE} can be approximately solved by invoking the separation of variables~\cite{Espinosa:2007qp}. Let us decompose the probability distribution into the time-dependent and field-dependent parts, $P(h,S,N) = \Psi(N) \Phi(h,S)$. Then, we have $\Psi \propto e^{-\alpha N}$ where $\alpha$ is constant to be determined by boundary conditions. Assuming the quantum fluctuation represented by the second term in the square brackets in \eqref{eq:FPE} dominates over the classical motion, we obtain
\begin{equation}
\bigg( \frac{\partial^2}{\partial h^2} + \frac{\partial^2}{\partial S^2} \bigg) \Phi 
= -\frac{8\pi^2}{H_{\rm inf}^2} \alpha \Phi \, .
\label{eq:Phi}
\end{equation}
Here we impose the boundary conditions adopted in~\cite{Hook:2014uia} in such a way that the probability vanishes where the classical motion $\delta\phi_{\rm cl} = -V_\phi/\big(3H_{\rm inf}^2\big)$  becomes larger than the quantum fluctuations $\delta \phi_{\rm qm} = H_{\rm inf}/(2\pi)$. If the classical motion becomes dominant in the unstable region, the Higgs quickly rolls down to the true vacuum with large negative potential and the space-time in such a region turns to the anti de Sitter space, leading to a crunch. On the other hand, even in a stable region, the classical force pulls the field back immediately after the quantum force kicks the field outside the region with $\delta \phi_{\rm qm} > \delta \phi_{\rm cl}$. Thus, the field never takes a value beyond the boundary.
Explicitly, we set the boundary conditions as 
\begin{equation}
P(\Lambda_h,S,N) = P(h,\Lambda_S,N) = 0 
\quad \text{with} \quad 
\bigg| \frac{\partial V}{\partial h} \bigg|_{h=\Lambda_h} 
=  \bigg| \frac{\partial V}{\partial S}\bigg|_{S=\Lambda_S} 
= \frac{3H_{\rm inf}^3}{2\pi} \, .
\label{eq:boundary}
\end{equation}
Taking into account the above boundary conditions and the axisymmetric property of the probability distribution $P(h,S,N) = P(-h,S,N) = P(h,-S,N)$, the general solution of \eqref{eq:Phi} is a superposition of sinusoidal mode functions:
\begin{equation}
\Phi = \sum_{n,m\geq0} A_{nm} \cos\left[\pi\bigg(n+\frac{1}{2}\bigg)\frac{h}{\Lambda_h} \right] 
\cos\left[\pi\bigg(m+\frac{1}{2}\bigg)\frac{S}{\Lambda_S} \right] \, .
\end{equation}
Then, $\alpha$ is parametrized by the integers $n$ and $m$ as
\begin{equation}
\alpha_{nm} = \bigg(n+\frac{1}{2}\bigg)^2\frac{H_{\rm inf}^2}{8\Lambda_h^2}
+ \bigg(m+\frac{1}{2}\bigg)^2\frac{H_{\rm inf}^2}{8\Lambda_S^2} \, . 
\end{equation}
$A_{nm}$ can be determined by the initial distribution. Substituting $P(h,S,0) = \delta(h)\delta(S)$, we obtain $A_{nm} = 1/(\Lambda_h \Lambda_S)$ and hence
\begin{equation}
P(h,S,N) = \frac{1}{\Lambda_h \Lambda_S} \sum_{n,m\geq0} e^{-\alpha_{nm} N} 
\cos\left[\pi\bigg(n+\frac{1}{2}\bigg)\frac{h}{\Lambda_h} \right] 
\cos\left[\pi\bigg(m+\frac{1}{2}\bigg)\frac{S}{\Lambda_S} \right] \, .
\label{eq:prob_dist}
\end{equation}

Let us consider the probability to realize our metastable universe at the end of inflation. The survival probability is calculated as
\begin{equation}
P_\Lambda = \int^{\Lambda_S}_{-\Lambda_S} dS 
\int^{\Lambda_{\rm max}}_{-\Lambda_{\rm max}} dh P(h,S) \, ,
\label{eq:survival_prob_def}
\end{equation}
where $\Lambda_{\rm max} = \Lambda_{\rm max}(S)$ is the critical value of $h$ where the Higgs potential is maximized. From $\partial V/\partial h(\Lambda_{\rm max},S) = 0$, we can obtain
\begin{equation}
\Lambda_{\rm max} = \sqrt{\frac{\lambda_{hS}}{-\partial\lambda_{\rm eff}/4\partial\log{h} - \lambda_{\rm eff}}}S 
\approx 10 \sqrt{\lambda_{hS}} S \, ,
\label{eq:Lambda_max}
\end{equation}
where for the second approximate equality we have used $-\partial\lambda_{\rm eff}/4\partial\log{h} - \lambda_{\rm eff} \approx -0.01$ which is valid for $h$ much larger than the conventional instability scale $\Lambda \sim 10^{10}$ -- $10^{11}$ GeV.
By using the analytic solution \eqref{eq:prob_dist}, we can easily integrate the survival probability \eqref{eq:survival_prob_def}, yielding 
\begin{equation}
P_\Lambda = \frac{4p}{\pi^2} \sum_{n,m\geq0} e^{-q\alpha_{nm}N} 
\frac{-\dfrac{m+1/2}{n+1/2} \Lambda_h^2 \sin\left[\beta \pi \left( n+\dfrac{1}{2} \right) \dfrac{\Lambda_S}{\Lambda_h}\right]+\beta \Lambda_h \Lambda_S}
{\beta^2 (n+1/2)^2\Lambda_S^2-(m+1/2)^2 \Lambda_h^2} \, ,
\label{eq:survival_prob}
\end{equation}
where $\beta \equiv \Lambda_{\rm max}/S$, and $p$ and $q$ are numerical fudge factors both of which are 1 in the above analytic estimation, but we adjust them to fit to the numerical results.

%%%%%%%%%%%%%%% MULTI-FIGURE  %%%%%%%%%%%%%%%
\begin{figure}[tp]
\centering
\subfigure[$\lambda_{hS} = 0.1$]{
\includegraphics [width = 7.5cm, clip]{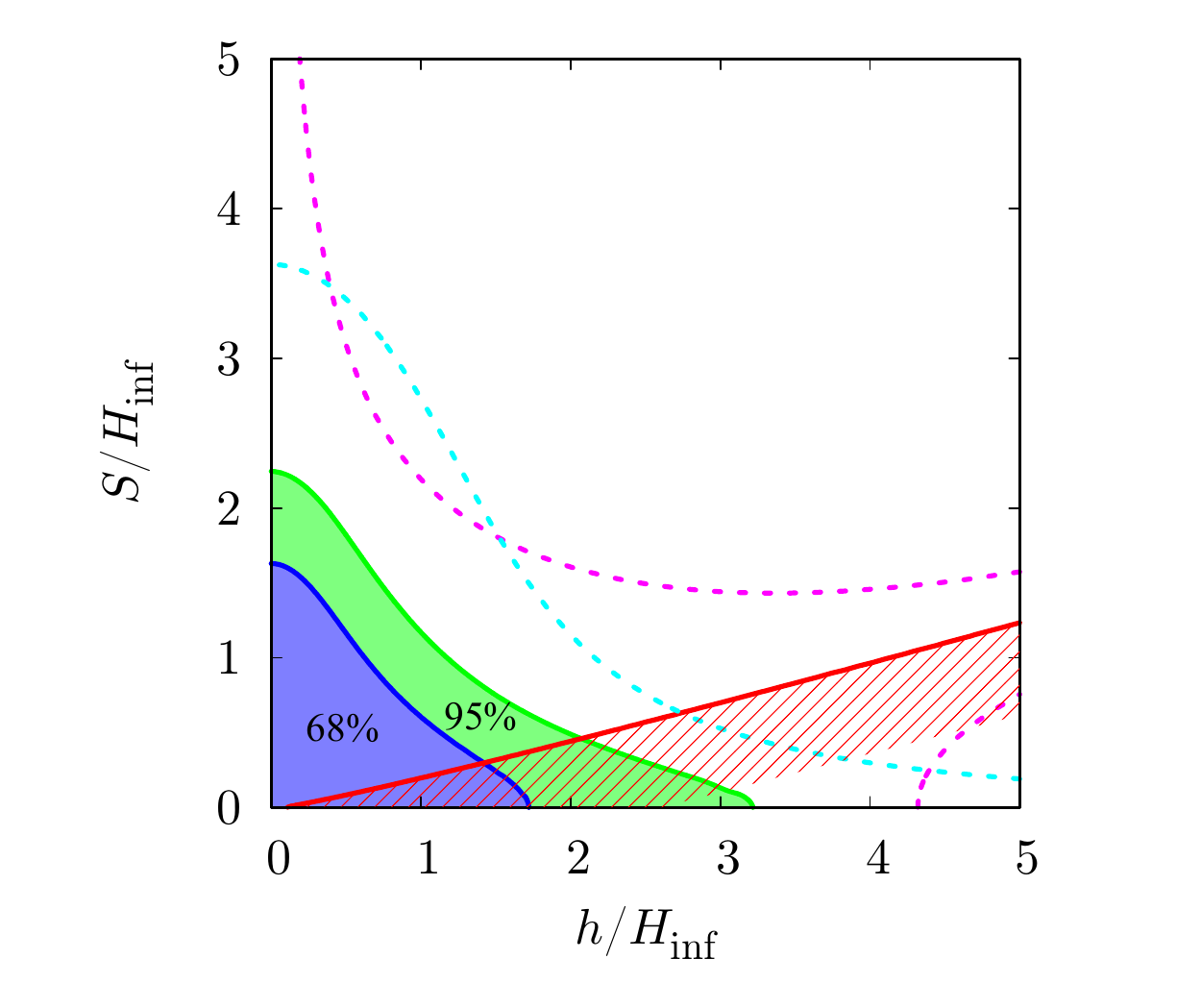}
\label{subfig:contour1}
}
\subfigure[$\lambda_{hS} = 10^{-4}$]{
\includegraphics [width = 7.5cm, clip]{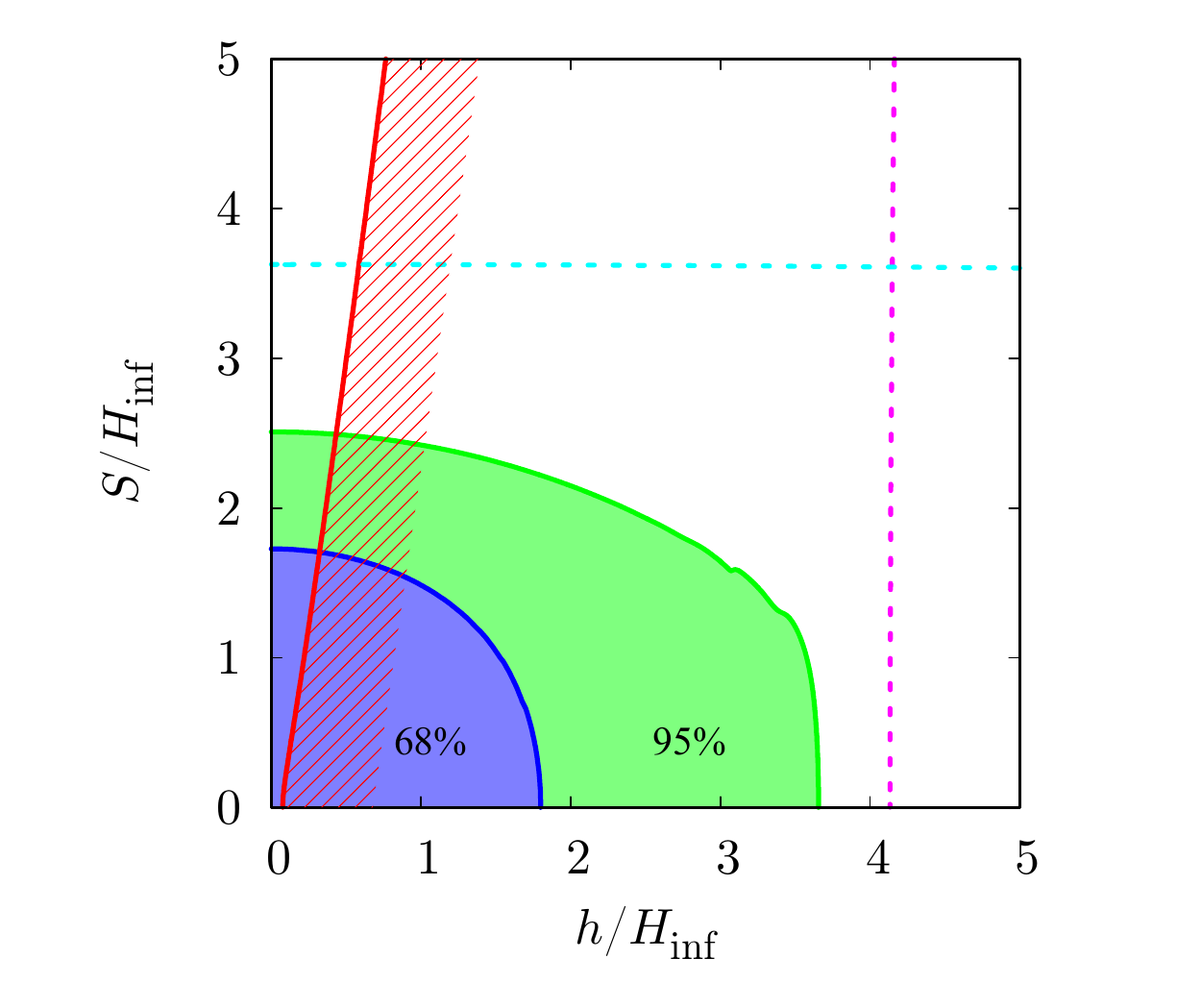}
\label{subfig:contour2}
}
\caption{
Equal probability contours corresponding to 68\% (blue) and 95\% (green) inside them. We have taken $H_{\rm inf} = 10^{11}$ GeV, $N = 60$, $\lambda_S = 0.01$ and $\lambda_{hS} = 0.1$ ($10^{-4}$) in the left (right) panel. The solid red lines represent the local maximum of the potential and the right side of the hatched region corresponds to the unstable region. The magenta and cyan dotted lines correspond to the boundary for $h$ and $S$ respectively beyond which $\delta\phi_{\rm cl} > \delta\phi_{\rm qm}$ so that stochastic approach is no longer valid.
}
\label{fig:contour}
\end{figure}
%%%%%%%%%%%%%%%%%%%%%%%%%%%%%%%%%%%%%%%

\begin{figure}[tp]
\centering
\includegraphics [width = 9cm, clip]{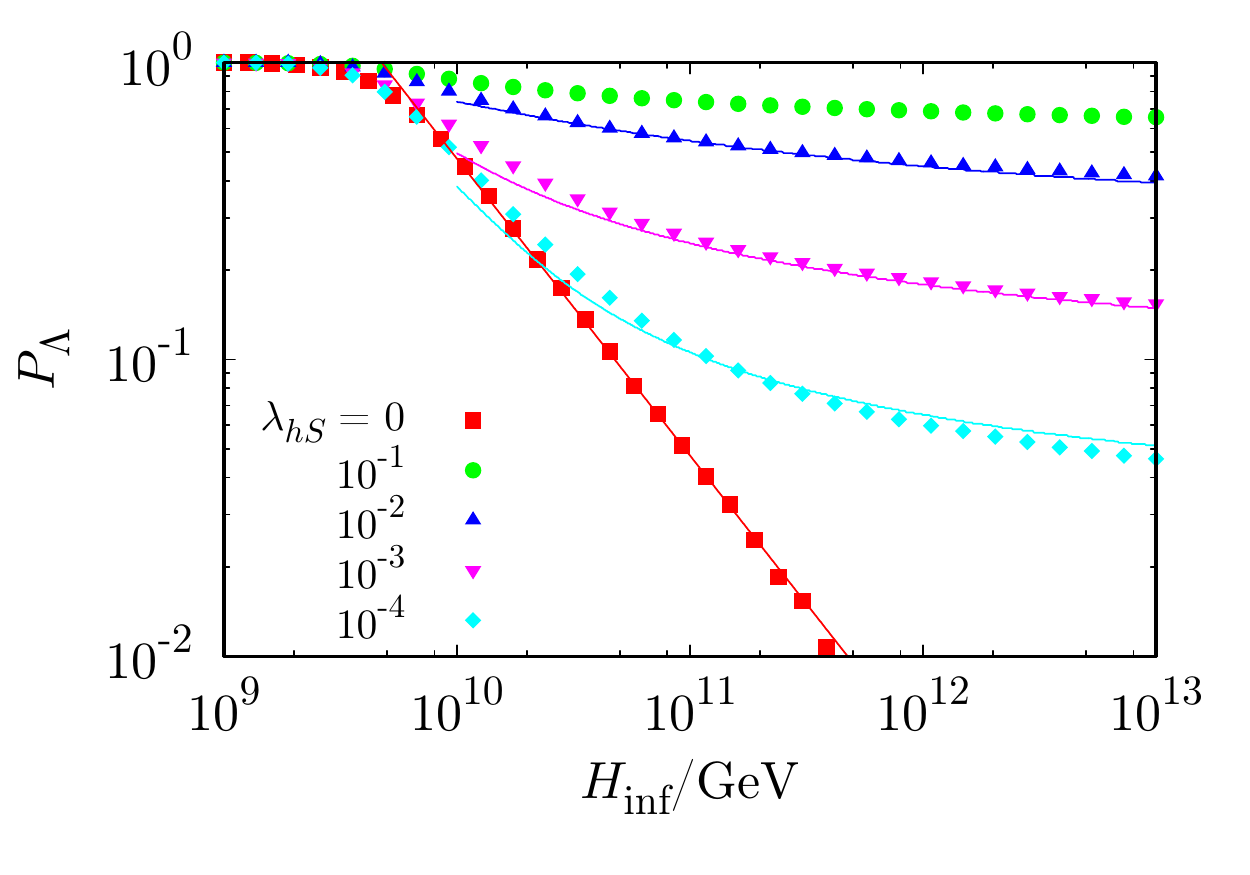}
\caption{
Survival probability in terms of the inflationary Hubble parameter. We have taken $N=60$, $\lambda_S = 0.01$ and $\lambda_{hS} = 0,\,10^{-4},\,10^{-3},\,10^{-2},\,10^{-1}$ from bottom to top. Each point and line correspond respectively to the numerical results and the analytic formula \eqref{eq:survival_prob} with $p=2$ and $q=5$.
}
\label{fig:survival_prob}
\end{figure}

We have numerically solved the Fokker-Planck equation \eqref{eq:FPE} under the same initial conditions and boundary conditions. Figure~\ref{fig:contour} shows contours of equal probability in $h$--$S$ plane. For large $\lambda_{hS}$, the probability distribution is deformed in such a way that both $h$ and $S$ cannot have large values simultaneously. More importantly, the critical value of $h$ maximizing the Higgs potential is proportional to $S$ as shown in \eqref{eq:Lambda_max}. It makes the electroweak vacuum stabilization more likely even in a case with small $\lambda_{hS}$. Figure~\ref{fig:survival_prob} shows the survival probability at $N=60$ from the beginning of inflation. It shows a significant enhancement of the survival probability even with small $\lambda_{hS}$. We have found that the analytic solution \eqref{eq:survival_prob} shows a good agreement with numerical results for $p=2$ and $q=5$.

%%%%%%%%%%%%%%% FIGURE  %%%%%%%%%%%%%%%
%%%%%%%%%%%%%%%%%%%%%%%%%%%%%%%%%%%

\section{Post-inflationary evolutions}

Let us consider the survival probability taking into account the post-inflationary evolution. Soon after the end of  inflation, the equations of motion for the Higgs and the singlet in a flat Friedmann universe are given by
\begin{align}
\ddot{h} + 3H\dot{h} - \frac{\Delta}{a^2}h + 
\left( \lambda_{\rm eff} + \frac{1}{4}\frac{\partial \lambda_{\rm eff}}{\partial \log h} \right) h^3 
+ \lambda_{hS} S^2 h &= 0 \, ,
\label{eq:eom_h} 
\\
\ddot{S} + 3H\dot{S} - \frac{\Delta}{a^2}S + \lambda_S S^3 + \lambda_{hS} h^2 S & = 0 \, .
\end{align}
For computational convenience, we approximate the value inside the parenthesis in \eqref{eq:eom_h} as constant by 
\begin{equation}
\lambda_{\rm eff} + \frac{1}{4}\frac{\partial \lambda_{\rm eff}}{\partial \log h} 
\equiv \tilde{\lambda}_{\rm eff} = 
\begin{cases} 
\lambda \quad & \text{for} \quad h < \Lambda 
\\[1mm] 
-\lambda \quad & \text{for} \quad h > \Lambda 
\end{cases}
\, ,
\end{equation}
with $\lambda = 0.01$ and $\Lambda = 10^{10}$ GeV. Note that as long as $\tilde{\lambda}_{\rm eff}$ takes a positive value, the amplitude of each field decreases inversely proportional to the scale factor.

%%%%%%%%%%%%%%% MULTI-FIGURE  %%%%%%%%%%%%%%%
\begin{figure}[tp]
\centering
\subfigure[$\lambda_{hS} = 0.1$]{
\includegraphics [width = 7.0cm, clip]{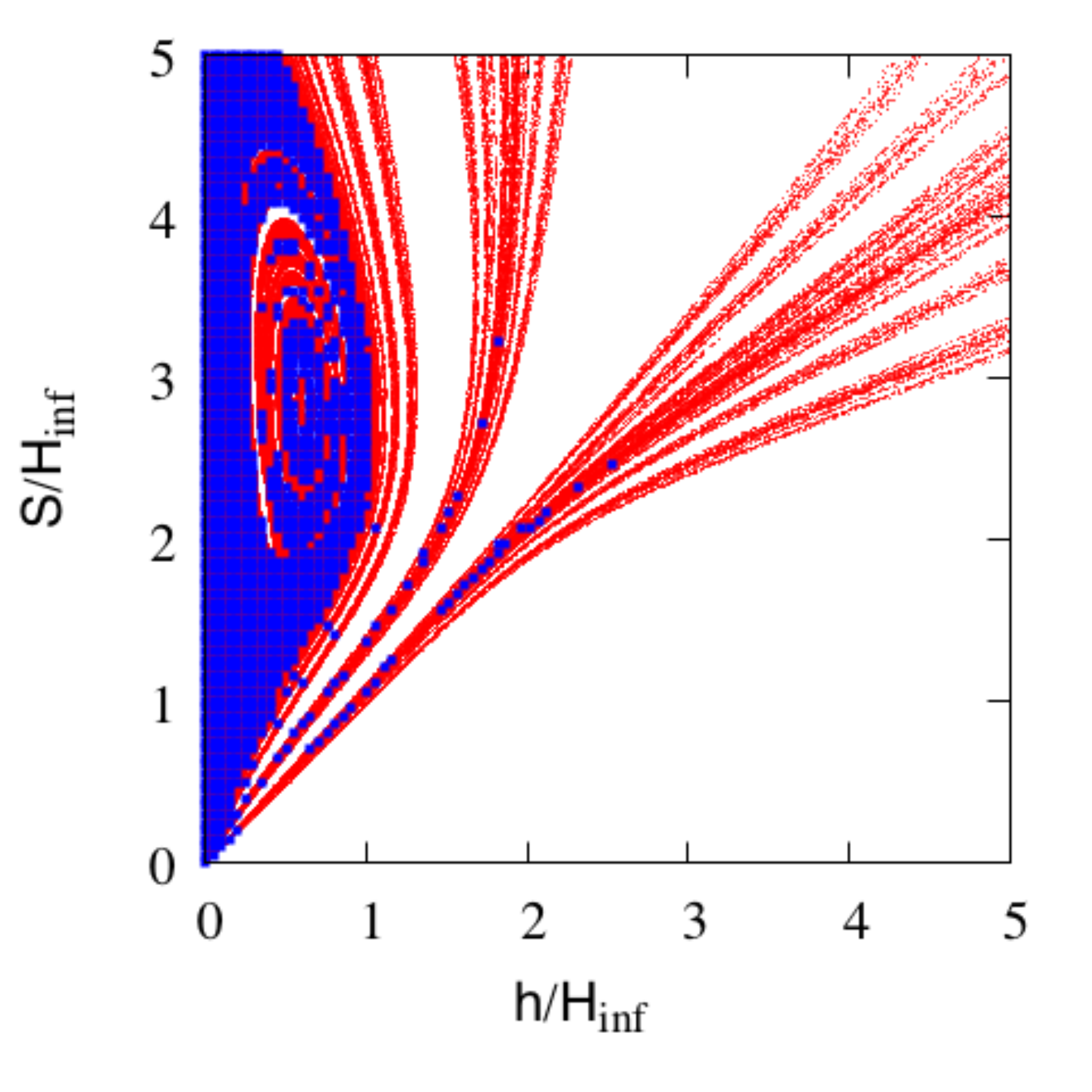}
\label{subfig:stability1}
}
\subfigure[$\lambda_{hS} = 0.03$]{
\includegraphics [width = 7.0cm, clip]{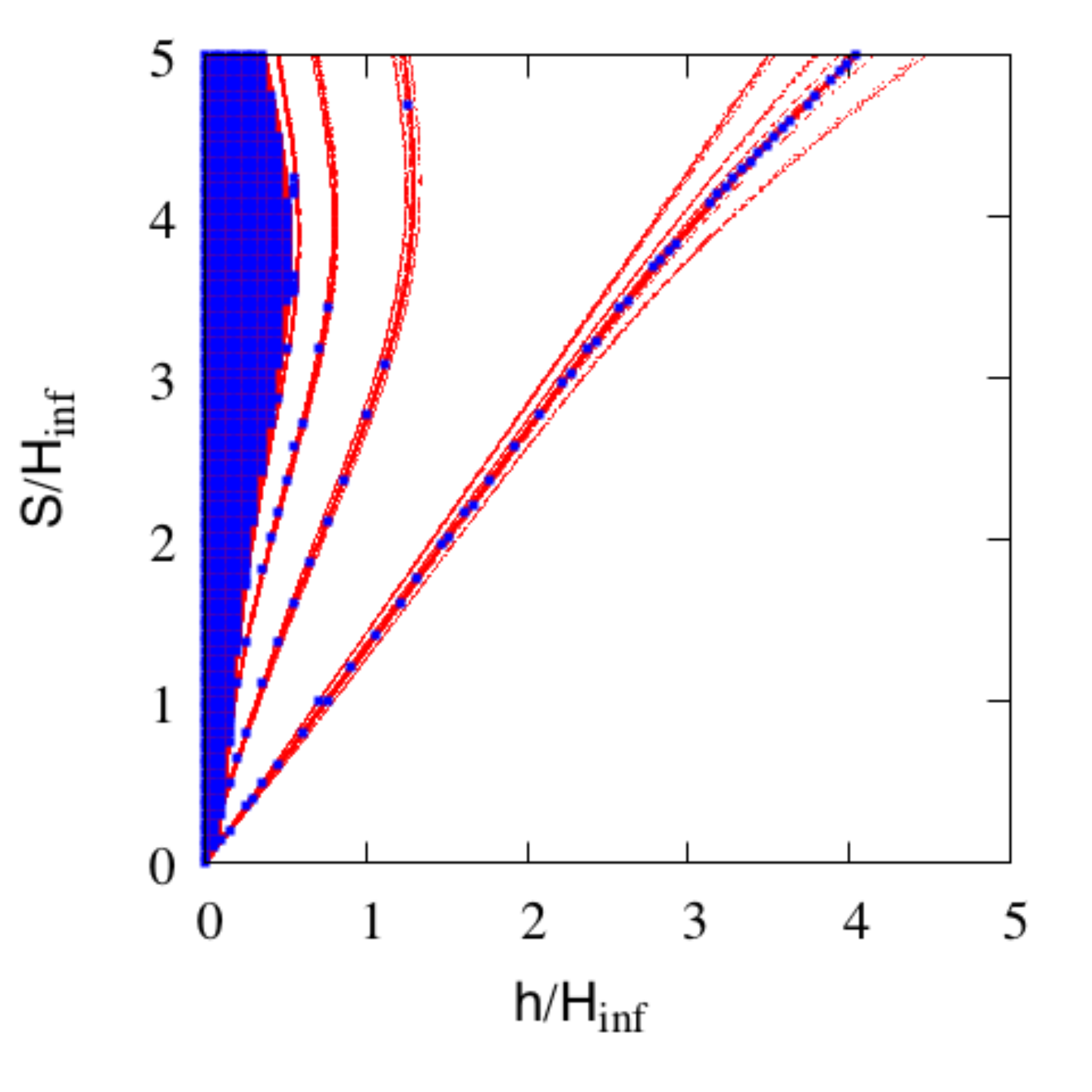}
\label{subfig:stability2}
}
\caption{
Post-inflationary stability chart for the Higgs from the numerical calculation of the homogeneous mode. The colored region corresponds to our metastable universe. We have taken $H_{\rm inf} = 10^{12}$ GeV, $\lambda_S = 0.01$ and $\lambda_{hS} = 0.1~(0.03)$ in the left (right) panel. The blue dots are the results from two-dimensional lattice computations.
}
\label{fig:stability}
\end{figure}
%%%%%%%%%%%%%%%%%%%%%%%%%%%%%%%%%%%%%%%

Because we assume $m_S \ll H_{\rm inf}$ so that the theory is nearly conformally invariant, parametric resonance occurs efficiently~\cite{Kofman:1997yn} in such a way that the oscillations of the singlet can cause the Higgs instability when it crosses zero, leading to the catastrophic destabilization even if the Higgs was stable before. We have numerically analyzed the post-inflationary dynamics in the above setup assuming matter domination with $H=2/(3t)$, and obtained a stability chart as shown in Figure~\ref{fig:stability}. The horizontal and the vertical axis correspond to the initial values of $h$ and $S$ just after inflation. We have found that for the zero modes the red region is stable, which is also supported by the two-dimensional lattice computations denoted by the blue dots, for each of which we have initially added a Gaussian random noise with the standard deviation $H_{\rm inf}/(2\pi)$.

\begin{figure}[tp]
\centering
\includegraphics [width = 9cm, clip]{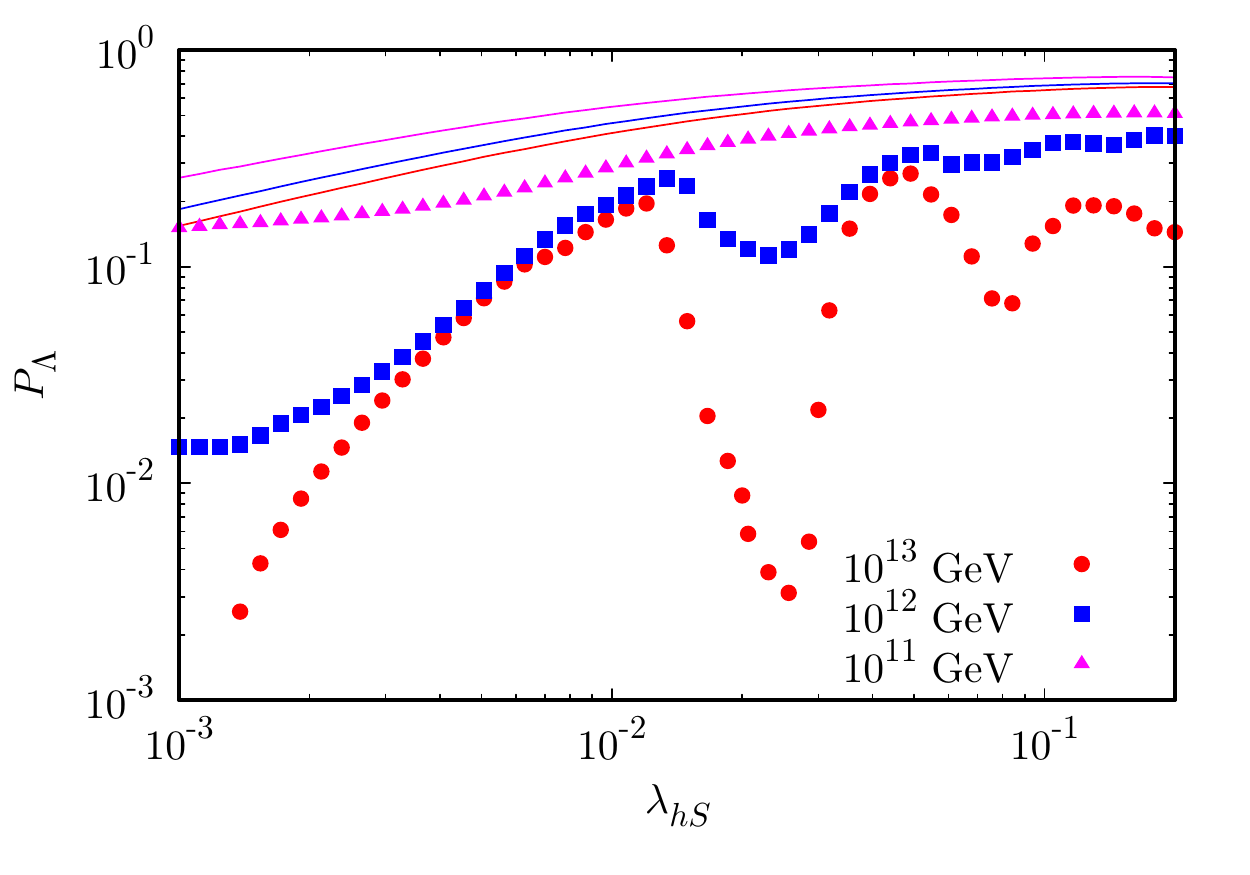}
\caption{
Survival probability well after the end of inflation in terms of $\lambda_{hS}$. We have taken $\lambda_S = 0.01$ and $H_{\rm inf} = 10^{13}$ GeV (circle), $10^{12}$ GeV (square) and $10^{11}$ GeV (triangle). Each solid line with same color corresponds to the probability at the end of inflation without destabilization due to parametric resonance.
}
\label{fig:survival_prob_post_inf}
\end{figure}

The survival probability taking into account the instability caused by parametric resonance between zero modes after inflation is shown in Figure~\ref{fig:survival_prob_post_inf}. As can be seen, the probability exhibits oscillations for different values of $\lambda_{hS}$, corresponding to the stability and instability bands: when $\lambda_{hS}/\tilde\lambda_{\rm eff}$ falls into the instability bands of the Lame equation~\cite{Greene:1997fu}, parametric resonance occurs for $h$ so that it may leave the stable region. The survival probability decreases for such values of $\lambda_{hS}$. Away from such $\lambda_{hS}$, however, still the electroweak vacuum has a good chance of stabilization, even for $H_{\rm inf} = 10^{13}$ GeV with probability $\sim 0.1$.

%%%%%%%%%%%%%%% FIGURE  %%%%%%%%%%%%%%%
%%%%%%%%%%%%%%%%%%%%%%%%%%%%%%%%%%%

\section{Discussions and conclusions}

We have studied a novel aspect for stabilizing the Higgs during inflation. With a singlet scalar coupled to the Higgs, the stochastic dynamics naturally gives rise to an enhanced probability of stable electroweak vacuum after inflation. In addition, the electroweak vacuum can survive the parametric resonance after inflation without suppressing the coupling. Furthermore, if the singlet is stable protected by Z$_2$ symmetry as we have discussed in this article, it can also take a role of present cold dark matter component \cite{McDonald:1993ex,Burgess:2000yq}. Although there is a stringent constraint on such a Higgs portal dark matter from collider and direct detection experiments, there still exist viable windows, $m_S \gtrsim 500$ GeV and $\lambda_{hS} \gtrsim 0.1$ or $m_S \sim m_H/2$ and $\lambda_{hS} \sim 10^{-3}$ -- 0.1 \cite{Cline:2013gha,He:2016mls}. We have seen that, in such a parameter region, the probability for stabilizing the electroweak vacuum can be significantly enhanced even for large $H_{\rm inf}$ compared with the conventional case, which means that the Higgs portal dark matter scenario is compatible with high-scale inflation scenarios.

Our scenario is attractive two-fold. First, other than the inflaton sector which we do not specify, we have only introduced a singlet scalar which can serve as dark matter, and thus our approach is minimal yet very effective. Furthermore, without invoking contrived potential or non-renormalizable interactions, the Higgs can stochastically land in the electroweak vacuum, so our scenario offers a natural solution to the Higgs stability via its cosmological evolution, especially for the high-scale inflation models. Investigating post-inflationary thermal history including the interaction between the Higgs and other standard model particles should give more information on the cosmological history of the Higgs.

\subsection*{Acknowledgments}

JG acknowledges the support from the Korea Ministry of Education, Science and Technology, Gyeongsangbuk-Do and Pohang City for Independent Junior Research Groups at the Asia Pacific Center for Theoretical Physics. JG is also supported in part by a TJ Park Science Fellowship of POSCO TJ Park Foundation and the Basic Science Research Program through the National Research Foundation of Korea Research Grant 2016R1D1A1B03930408. 
NK acknowledges the support by Grant-in-Aid for Japan Society for the Promotion of Science Fellows.

\end{document}